\documentclass[nofootinbib,onecolumn,amsmath,showpacs,12pt]{revtex4-2}
\usepackage{graphicx}
\usepackage{dcolumn}
\usepackage{bm}
\usepackage{color} 
\usepackage{slashed}
\begin{document}
\newcommand{\hs}{\hspace*{0.5cm}}
\newcommand{\vs}{\vspace*{0.5cm}}
\newcommand{\be}{\begin{equation}}
\newcommand{\ee}{\end{equation}}
\newcommand{\bea}{\begin{eqnarray}}
\newcommand{\eea}{\end{eqnarray}}
\newcommand{\ben}{\begin{enumerate}}
\newcommand{\een}{\end{enumerate}}
\newcommand{\bde}{\begin{widetext}}
\newcommand{\ede}{\end{widetext}}
\newcommand{\nn}{\nonumber}
\newcommand{\crn}{\nonumber \\}
\newcommand{\Tr}{\mathrm{Tr}}
\newcommand{\non}{\nonumber}
\newcommand{\noi}{\noindent}
\newcommand{\al}{\alpha}
\newcommand{\la}{\lambda}
\newcommand{\bet}{\beta}
\newcommand{\ga}{\gamma}
\newcommand{\va}{\varphi}
\newcommand{\om}{\omega}
\newcommand{\pa}{\partial}
\newcommand{\+}{\dagger}
\newcommand{\fr}{\frac}
\newcommand{\bc}{\begin{center}}
\newcommand{\ec}{\end{center}}
\newcommand{\Ga}{\Gamma}
\newcommand{\de}{\delta}
\newcommand{\De}{\Delta}
\newcommand{\ep}{\epsilon}
\newcommand{\varep}{\varepsilon}
\newcommand{\ka}{\kappa}
\newcommand{\La}{\Lambda}
\newcommand{\si}{\sigma}
\newcommand{\Si}{\Sigma}
\newcommand{\ta}{\tau}
\newcommand{\up}{\upsilon}
\newcommand{\Up}{\Upsilon}
\newcommand{\ze}{\zeta}
\newcommand{\ps}{\psi}
\newcommand{\Ps}{\Psi}
\newcommand{\ph}{\phi}
\newcommand{\vph}{\varphi}
\newcommand{\Ph}{\Phi}
\newcommand{\Om}{\Omega}
\newcommand{\AdrHEPC}{Phenikaa Institute for Advanced Study, Phenikaa University, Nguyen Trac, Duong Noi, Hanoi 100000, Vietnam}

\title{Quasi-Dirac fermion: A source of neutrino mass and dark matter} 

\author{Nguyen Thi Nguyet Nga}
\email{nganguyet4988@hvu.edu.vn}
\affiliation{Faculty of Natural Sciences, Hung Vuong University, Nong Trang, Phu Tho, Vietnam}

\author{Nguyen Huy Thao}
\email{nguyenhuythao@hpu2.edu.vn}
\affiliation{Faculty of Physics, Hanoi Pedagogical University 2, Xuan Hoa, Phu Tho, Vietnam}

\author{Phung Van Dong} 
\email{dong.phungvan@phenikaa-uni.edu.vn (corresponding author)}
\affiliation{\AdrHEPC} 

\date{\today}

\begin{abstract}

Neutral vectorlike fermion as inspired by unified theories might become quasi-Dirac states at TeV due to a violation in lepton-like symmetry. It is shown that such quasi-Dirac fermions can properly achieve radiative neutrino mass generation and dark matter stability. Indeed, the small splitting of quasi-Dirac masses, i.e. $\Delta M/M\ll 1$, suitably suppresses neutrino mass to be small in order to allow dark matter annihilation and detection to be appropriate to experiment as well as charged lepton flavor violation limit. 
                  
\end{abstract} 

\maketitle

\section{Introduction}

Neutrino mass \cite{kajita,mcdonald} and dark matter \cite{bertone,arcadi} are the important issues in the modern physics, dictating that the standard model must be extended. The simplest scheme of which includes a right-handed neutrino $\nu_R$ to each family as associated with usual left-handed neutrino $\nu_L$. The relevant neutrino mass is generated by a seesaw mechanism \cite{minkowski,yanagida,gell-mann,mohapatra,valle}, i.e. $m_\nu\simeq -\fr{m^2}{M}$, where $m=-h v/\sqrt{2}$ for $v=246$ GeV is a Dirac mass that couples $\nu_L$ to $\nu_R$, while $M$ is a Majorana mass that couples $\nu_R\nu_R$ by themselves. Since the neutrino mass is fixed, i.e. $m_\nu\sim 0.1$ eV \cite{pdg}, the new physics mass $M$ scales as $a^2$, given that the Dirac coupling $h$ scales as $a$. The seesaw mechanism matches the GUT prediction for a unified coupling $h\sim 0.57$ (in conventional notation, $h^2/4\pi\sim 40^{-1}$) at energy scale $M\sim 10^{14}$ GeV, hence it is potentially motivated by a GUT \cite{gut1,gut2,gut3,gut4,gut5,gut6}. However, an intermediate new physics responsible for dark matter expected at energy scale $M\sim$ TeV requires $h\sim 1.8\times 10^{-6}$ for explaining the neutrino mass, which is unlikely. Indeed, this prediction of the $h$ coupling is not allowed by the RGE where the inverse square coupling changes as logarithm of relative energy scales, i.e. only in one order of magnitude as of $h\sim 0.05$, by contrast.\footnote{Indeed, the running of $h$ depends on the Higgs self-coupling besides others, which is interrupted at a middle energy around $10^{10}$ GeV, where the Higgs vacuum becomes unstable. On the other hand, a small coupling such as $h\lesssim 10^{-3}$ does not affect the running of the Higgs self-coupling below and around this regime. Given a connection of $h$ before and after this regime relevant to the Higgs vacuum stability, one would impose $h$ to be sizable below the regime, for which $h\sim 0.05$ is taken \cite{ssrc,ssrc1}.} Furthermore, the seesaw mechanism in itself does not explain the issue of dark matter stability. 

Nontrivial generalization of the seesaw mechanism introduces a $Z_2$ symmetry and an extra Higgs doublet $\eta$, besides $\nu_R$, such that both $\eta$ and $\nu_R$ are odd under $Z_2$, while the standard model fields are even under this group, called scotogenic mechanism \cite{tao,ma}. The neutrino mass is now radiatively induced to be $m_\nu\sim (1/16\pi^2)(\la h^2)(v^2/M)$, where $M$ is a typical large mass of $\Re(\eta^0),\Im(\eta^0),\nu_R$ running in the loop, which is expected at TeV, $\la$ is the coupling of $\eta$ and usual Higgs doublet, and $h$ is now the coupling of $\nu_L$ to $\eta^0$ and $\nu_R$. The neutrino mass is thus suppressed by the loop factor $1/16\pi^2\sim 6.3\times 10^{-3}$, in addition to $v/M$ as of the seesaw. This case acquires $\la h^2\sim 2.6\times 10^{-10}$, or roundly $\la \sim 10^{-4}$ and $h\sim 10^{-3}$, which are significantly bigger than that in the seesaw. Additionally, a direct result of this setup is that the lightest of $\Re(\eta^0)$, $\Im(\eta^0)$, and $\nu_R$ is stabilized by $Z_2$, responsible for dark matter. However, the predictions of $\la,h$ are still small in order to make such dark matter phenomenologically viable. Namely, \ben
\item The fermion dark matter candidate, i.e. $\nu_R$, overpopulates the universe, because of $\langle \sigma v\rangle_{\nu_R\nu_R \to \nu\nu,ee}\lesssim h^4 (\mathrm{TeV}/M)^2\ \mathrm{pb}\ll 1$ pb, even for $h\lesssim 0.05$ implied by the RGE, unless an unreasonable coannihilation between the dark matter $\nu_R$ and a dark scalar $\Re(\eta^0)$, $\Im(\eta^0)$, or $\eta^\pm$ due to a fine-tuning in the relevant masses occurs.\footnote{The benchmark $M\sim 1$ TeV originates from the subsequent charged-lepton flavor violation.}\item An alternative matter, the charged-lepton flavor violation decay derived by the coupling $\mathcal{L}\supset h \bar{l}_L \eta \nu_R +H.c.$ evaluated by $\mathrm{Br}(\mu \to e\ga)\lesssim 4.2\times 10^{-13} (h/0.05)^4 (\mathrm{TeV}/M)^4 \leq 4.2\times 10^{-13}$ requires $h\lesssim 0.05$, for $M\sim $ TeV, in agreement with the RGE, which also implies that $\nu_R$ cannot be dark matter in this kind of the model \cite{vicente}. 
\item The scalar dark matter candidate, i.e. either $\Re(\eta^0)$ or $\Im(\eta^0)$, is not realistic due to a large scattering with nuclei via $Z$ exchange, unless $\la$ is large enough to order to split these scalar masses which kinematically suppresses the process, say $\Delta M\simeq \la v^2/2M\geq 200$--300~MeV---the maximum transfer momentum of a TeV dark matter scattering off a nucleus with typical mass 100--150 GeV, which requires $\la\gtrsim 0.01$ \cite{msplt}.\een  

It is clear that only the loop suppression in the scotogenic neutrino mass does not naturally fit the dark matter observables and the other bounds. We would like to suggest in this work a nontrivial generalization of the scotogenic scheme by introducing a quasi-Dirac fermion $N_{L,R}$ instead of $\nu_R$. As a result, the neutrino mass is suppressed by the corresponding quasi-Dirac approximation, namely $\delta M/M\sim 10^{-5}$, in addition to that of the scotogenic mechanism, where $M$ is a Dirac mass at TeV scale connecting $N_L$ and $N_R$, while $\delta M\sim 10$ MeV summarizes small Majorana masses of $N_L$ and $N_R$ which make the Dirac (or vectorlike) state $N_{L,R}$ becomes quasi (or pseudo) $N_{1,2}$. In other words, this model predicts a radiative inverse-seesaw neutrino mass \cite{ins1,ins2,ins3}, whereas that of the scotogenic setup belongs to the form of a radiative canonical seesaw. Due to such quasi-Dirac approximation, i.e. nearly-degenerate masses, the model now acquires $\la h^2\sim 2.6\times 10^{-5}$. That said, the Yukawa coupling of dark fermion with usual neutrino may be sizable, comparable to the RGE prediction and the charged-lepton flavor violation limit, i.e. $h\sim 0.05$. Additionally, the scalar dark matter is now available under direct detection due to $\la\sim 0.01$, while it obtains a correct abundance via the gauge and/or Higgs portal. 

In what follows, we first present this scheme and determine generated neutrino masses. We then examine dark matter observables and related constraints. We finally discuss extensions that are inspired by this proposal and conclude this work.                    

\section{Quasi-Dirac scheme}

We introduce to the standard model an extra Higgs doublet $\eta$ and a new fermion $N$ for each family so that $\eta$ and $N$ are odd under a $Z_2$ symmetry. Hereafter, the family index is suppressed, unless necessary. The particle representation content under the relevant symmetries is summarized in Table \ref{tab1}.

\begin{table}[h]
\begin{tabular}{lcccc}
\hline\hline
Field & $SU(3)_C$ & $SU(2)_L$ & $U(1)_Y$ & $Z_2$ \\
\hline
$l_L=\begin{pmatrix}
\nu_L\\
e_L\end{pmatrix}$ & 1 & 2 & $-\fr{1}{2}$ & +\\
$q_L=\begin{pmatrix}
u_L\\
d_L\end{pmatrix}$ & 3 & 2 & $\fr{1}{6}$ & +\\
$e_R$ & 1& 1& $-1$ & +\\
$u_R$ & 3 & 1 & $\fr{2}{3}$ & +\\
$d_R$ & 3 & 1 & $-\fr{1}{3}$ & +\\
$N_{L,R}$ & 1 & 1 & 0 & $-$\\
$\phi=\begin{pmatrix}
\phi^+\\
\phi^0\end{pmatrix}$ & 1 & 2 & $\fr{1}{2}$ & +\\
$\eta=\begin{pmatrix}
\eta^0\\
\eta^-\end{pmatrix}$ & 1 & 2 & $-\fr{1}{2}$ & $-$ \\
\hline\hline
\end{tabular}
\caption[]{\label{tab1} Particle representation content of the model.}
\end{table} 

Lagrangian of the model takes the form, 
\be \mathcal{L} = \mathcal{L}_{\mathrm{kin}}+\mathcal{L}_{\mathrm{Yuk}}-V(\phi,\eta),\ee where $\mathcal{L}_{\mathrm{kin}}$ summarizes over the kinetic terms of all fields, while the Yukawa interactions and the scalar potential are given, respectively, by 
\bea \mathcal{L}_{\mathrm{Yuk}} &=& h^e \bar{l}_L \phi e_R + h^d \bar{q}_L \phi d_R + h^u \bar{q}_L \tilde{\phi} u_R + h \bar{l}_L \eta N_R  + h' \bar{l}_L \eta N^c_L \crn && - M \bar{N}_L N_R -\fr 1 2 \mu_L \bar{N}_L N^c_L -\fr 1 2\mu_R \bar{N}^c_R N_R + H.c.\eea
\bea V(\phi,\eta) &=& \mu^2_1\phi^\dagger \phi +\mu^2_2 \eta^\dagger \eta + \la_1 (\phi^\dagger \phi)^2+\la_2 (\eta^\dagger \eta)^2\crn
&&+ \la_3 (\phi^\dagger \phi)(\eta^\dagger \eta)+\la_4 (\phi^\dagger \eta)(\eta^\dagger \phi)+\fr 1 2 \la_5[(\phi \eta)^2+H.c.]\eea Above, the couplings $\la$'s and $h$'s are dimensionless, while the parameters $\mu$'s and $M$ have a mass dimension. Additionally, we can choose a basis so that $\la_5$ is real as factorized out. 

Since $M$ is a Dirac mass conserving every symmetry, it can be as large as the cut-off scale, whereas $\mu_{L,R}$ and $h'$ would be small. Indeed, in the limit $\mu_{L,R}\to 0$ and $h'\to 0$, the theory conserves a lepton-like symmetry, such as $f\to e^{i\al}f$ for $f=l_L,e_R, N_{L,R}$. Hence, the smallness of $\mu_{L,R}$ and $h'$ is due to this symmetry protection, i.e. naturally explained by a more fundamental theory via a large scale or loops. That said, $\mu_{L,R}\ll M$ and $h'\ll h$. It is suitably imposed $h'/h\sim \mu_{L,R}/M\ll 1$, since these ratios vanish for the conservation of the lepton-like symmetry. Further, the new fermion $N$ gains a mass Lagrangian,
\be \mathcal{L}_{\mathrm{Yuk}}\supset -\fr 1 2 \begin{pmatrix} \bar{N}_L & \bar{N}^c_R \end{pmatrix}\begin{pmatrix}
\mu_L & M \\
M & \mu_R\end{pmatrix}\begin{pmatrix}
N^c_L\\
N_R\end{pmatrix}.\ee Because of $\mu_{L,R}\ll M$, the field $N$ is a quasi-Dirac fermion, related to that in mass basis as
\be \begin{pmatrix} 
N^c_L\\
N_R\end{pmatrix}=U\begin{pmatrix} N_{1R} \\
N_{2R}\end{pmatrix},\hs U = \begin{pmatrix} c_\theta & s_\theta \\
-s_\theta & c_\theta
\end{pmatrix}, \ee where $\cot({2\theta}) = (\mu_R-\mu_L)/2M \ll 1$, or \be \theta \simeq \fr{\pi}{4}+\fr{\mu_L-\mu_R}{4M},\ee i.e. $s_\theta\simeq c_\theta \simeq 1/\sqrt{2}$ up to $\mu_{L,R}/M$ corrections. The physical fields $N_{1,2}$ obtain a mass, 
\be m_{N_1}\simeq -M+\fr 1 2 (\mu_L+\mu_R),\hs m_{N_2}\simeq M+\fr 1 2 (\mu_L+\mu_R).\ee This is exactly a quasi-Dirac approximation, as proposed, governing neutrino mass and dark matter, as shown below.      

Because of $Z_2$ symmetry, the field $\phi=[G^+_W,(v+H+iG_Z)/\sqrt{2}]$ behaves as the standard model Higgs doublet, i.e. $G^+_W=\phi^+$ and $G_Z=\sqrt{2}\Im(\phi^0)$ are the Goldstone bosons associated with gauge bosons, $W^+$ and $Z$, respectively, while $H=\sqrt{2}\Re(\phi^0)-v$ is the usual Higgs boson with mass $m_H=\sqrt{2\la_1}v$, where $v=246$ GeV is the weak scale, as usual. The dark field $\eta=[(S+iA)/\sqrt{2},H^-]$ in which $H^-=\eta^-$, $S=\sqrt{2}\Re(\eta^0)$, and $A=\sqrt{2}\Im(\eta^0)$ are physical fields with masses given, respectively, by
\bea m^2_{H^-} &=& \mu^2_2+(\la_3 +\la_4) v^2/2,\\
m^2_S&=&\mu^2_2+(\la_3+\la_5)v^2/2,\\ 
m^2_A&=&\mu^2_2+(\la_3-\la_5)v^2/2.\eea  

\begin{figure}[h]
\bc
\includegraphics[scale=0.8]{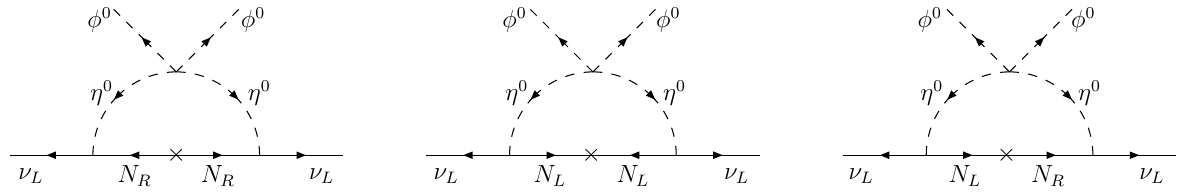}
\caption[]{\label{fig1} Radiative corrections to neutrino mass}
\ec
\end{figure}
Neutrino gains a mass from radiative corrections as depicted in Fig. 1, given in gauge basis. Hence, we derive 
\bea m_\nu &=& \fr{i}{2} \int \fr{d^4 p}{(2\pi)^4}\fr{(h U_{2I}+h'U_{1I})^2m_{N_I}(m^2_S-m^2_A)}{(p^2-m^2_{N_I})(p^2-m^2_S)(p^2-m^2_A)},
\eea which is summed over $I=1,2$. With the aid of the quasi-Dirac approximation, we have 
\bea m_\nu &\simeq & i \int \fr{d^4 p}{(2\pi)^4}\fr{h^2\mu M^2(m^2_S-m^2_A)}{(p^2-M^2)^2(p^2-m^2_S)(p^2-m^2_A)}\crn
&&+i \int \fr{d^4 p}{(2\pi)^4}\fr{h^2 \kappa (m^2_S-m^2_A)}{(p^2-M^2)(p^2-m^2_S)(p^2-m^2_A)},
\eea where $\mu=(\mu_L+\mu_R)/2$ and $\kappa=\mu_R/2+Mh'/h$. We get further
\bea m_\nu &\simeq& \fr{\la_5 h^2}{16\pi^2}\fr{v^2}{\bar{M}^2-M^2}\left\{\kappa\left(1-\fr{M^2\ln\fr{\bar{M}^2}{M^2}}{\bar{M}^2-M^2}\right)\right.\crn
&&\left.+\mu\left[\fr{2M^2}{\bar{M}^2-M^2}-\fr{M^2(\bar{M}^2+M^2)\ln\fr{\bar{M}^2}{M^2}}{(\bar{M}^2-M^2)^2}\right]\right\},\eea where $\bar{M}^2=(m^2_S+m^2_A)/2$. Noticing that $\kappa\sim \mu \ll M\sim \bar{M}$, the neutrino mass is suppressed by $(\mu,\kappa)/(M,\bar{M})$ compared to that induced by the usual scotogenic setup, i.e. 
\be m_\nu \sim \fr{\la_5 h^2}{16\pi^2}\fr{v^2}{(M,\bar{M})}\fr{(\mu,\kappa)}{(M,\bar{M})}.\label{dds}\ee That said, taking $(\mu,\kappa)/(M,\bar{M})\sim 10^{-5}$, it leads to $\la_5\sim 0.01$ and $h\sim 0.05$ in order to fit the neutrino oscillation data, as expected.

Remarks are in order. \ben 
\item Our neutrino mass generation scheme reveals a novel type of radiative inverse seesaw for which the radiative correction is directly related to neutrino mass, not to small Majorana masses $\mu,\kappa$ as of the normal scene \cite{rins1,rins2}. Additionally, while our scheme inherits a compelling feature as mass correction induced by scotogenic (i.e., dark) fields, it yields a radiative inverse-seesaw neutrino mass opposite to the scotogenic setup which generally gives a radiative canonical-seesaw neutrino mass, by contrast. That said, the dark fermions that couple to usual neutrinos may be dark matter, whereas those (i.e., vectorlike neutral fermions) in radiative inverse seesaw may decay to usual neutrinos and ordinary Higgs boson.
\item In the usual scotogenic setup, i.e. $N_L$ is omitted (thus, $\mu_L,M,h'$ are removed too), one can particularly impose $\mu_R$---the Majorana mass of $N_R$---to be small, i.e. $\mu_R\ll \bar{M}$ \cite{lsrn}. The resultant neutrino mass behaves as (\ref{dds}), but the dark matter candidate is only the light Majorana fermion $N_R$ and is generally ruled out by the relic density bound. In our model, the dark fermion $N_{L,R}$ (i.e., $N_{1,2}$) can have an arbitrarily-large mass. Additionally, the dark scalar $S,A$ can be a dark matter candidate if the lightest of them has a mass smaller than the quasi-Dirac masses of $N_{1,2}$. 
\een

Reference \cite{addref1} has generalized the minimal scotogenic scheme by including an arbitrary number of inert scalar doublets, say $\eta$'s, and another arbitrary number of right-handed fermion singlets, say $N$'s, in which all $\eta$'s and $N$'s are odd under a $Z_2$ parity. The work has discussed nontrivial contributions to neutrino mass coming from the inert doublets $\eta$'s with a possibility that their small couplings with usual Higgs fields ($\la_5$'s) and the $Z_2$ symmetry are maintained at high energy. However, the right-handed fermion singlets have been assumed to be flavor diagonal, i.e. they have Majorana masses by themselves, for which the contributions of such right-handed fermion singlets to neutrino mass are simply added up from each individual contribution, as in the minimal scotogenic setup for three right-handed fermion singlets. At the first look, our proposal simply includes three $N_R$'s and three $N^c_L$'s, which are just six right-handed fermion singlets as proposed in \cite{addref1}. However, the difference lies at global lepton number assignment, the original scotogenic model and its generalization in \cite{addref1} put $L(N)=0$ for all $N$'s and $L(\eta)=-1$ for all $\eta$'s, while our model assigns $L(N)=1$ for all $N$'s and $L(\eta)=0$ for the inert scalar doublet, by contrast, which we recast it as a lepton-like charge, as supplied above, since this new charge might explain the $Z_2$ parity by gauge completion, discussed at the end. Interestingly, our proposal gives rise to novel physical results, because of the interplay of lepton-like symmetry and of pure-Majarona versus quasi-Dirac mass question. First of all, the lepton-like charge conservation demands that the Yukawa couplings of $N^c_L$'s to usual neutrinos $\nu_L$'s which violate this charge are small, i.e. $h'\ll h$, as given above, which are different from those in \cite{addref1} for which both $h,h'$ conserve lepton number. Second, there is no symmetry reason that forbids $N$'s from obtaining a Dirac mass in both approaches, in our setup $N_{L,R}$ gain a Dirac mass as $M \bar{N}_L N_R$, while in previous setup~\cite{addref1}, the Dirac mass is $M \bar{N}_i N^c_j$, where $N^c_j$ can be identified as the left-handed partner of the right-handed fermion $N_i$, for $i\neq j$. That said, in the setup \cite{addref1}, the $\la_5$ couplings among inert scalars and usual Higgs fields were small due to the violation of lepton number, but the pure-Majorana masses, say $m_{N_i} \bar{N}_i N^c_i$, conserved this lepton number to be large, which unfortunately did not exclude the Dirac mass $M \bar{N}_i N^c_j$, by contradiction. In other words, the hypothesis of pure-Majorana masses for $N$'s is unreasonable in any case. In our setup, the masses of $N$'s are determined by the lepton-like symmetry, namely the pure-Majorana masses $\mu_L \bar{N}_L N^c_L$ and $\mu_R \bar{N}^c_R N_R$ which violate the lepton-like symmetry must be small, compared to the Dirac mass $M\bar{N}_L N_R$, that is $\mu_{L,R}\ll M$, as the Yukawa couplings are, i.e. $h'\ll h$, by the 't Hooft's naturalness \cite{addref2}. Here, the Dirac mass $M$ conserves lepton-like (even lepton) charge, which would be large. This gives rise to quasi-Dirac fermions in our setup. Intriguingly, the quasi-Dirac fermions give nearly degenerate (and opposite) Majorana masses, i.e. $M$ and $-M$, for which their contributions to neutrino mass are cancelled out. What results is a novel neutrino mass suppressed by $\mu_{L,R}/M$, as given above, which is a new observation of this work.                

\section{Dark matter and related phenomena}

\begin{figure}[h]
\bc
\includegraphics[scale=1]{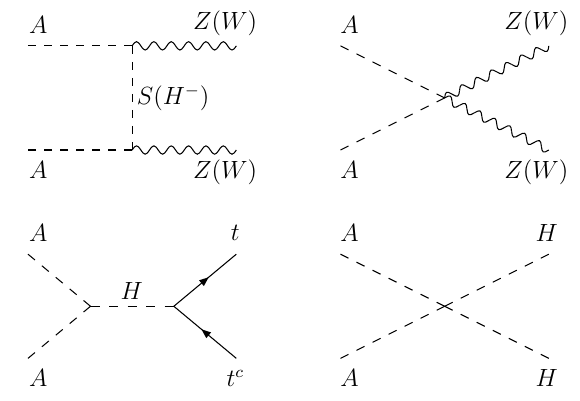}
\caption[]{\label{fig2} Annihilation processes of $A$'s that set the dark matter abundance.}
\ec
\end{figure}
This model predicts the existence of a scalar dark matter, which is assumed to be $A$, i.e. $m_A<m_{S}$, $m_A<m_{H^-}$, and $m_A<m_{N_{1,2}}$, implying $\la_5>\mathrm{Max}(-\la_4,0)$ and $\bar{M}<M$. In the early universe, $A$ annihilates to usual particles via the gauge and Higgs portals, as depicted in Fig. \ref{fig2}, where there would exist a $u$-channel diagram to $ZZ$ associated with $t$-channel diagram to $ZZ$, which is skipped for brevity. Further, the $s$-channel diagram by $H$ portal may produce other particle pairs of the standard model, but their contributions are radically smaller that of $tt^c$ and are skipped too. 

The processes are computed, based on the relevant couplings supplied as \bea \mathcal{L} &\supset&-\fr{m_t}{v}\bar{t}t H -\fr{\la_3-\la_5}{4}\left(2v H+H^2\right)A^2+\fr{g}{2c_W}Z^\mu S\stackrel{\leftrightarrow}{\pa}_\mu A\crn
&&+\left(\fr{g}{2}W^{+\mu} H^-\stackrel{\leftrightarrow}{\pa}_\mu A+H.c.\right)+\fr{g^2}{4}\left(\fr{Z^2}{2c^2_W}+W^+ W^-\right)A^2.\eea The annihilation cross section is thus evaluated by 
\be \langle \sigma v \rangle \simeq \fr{(\la_3-\la_5)^2 m^2_t}{64\pi m^4_A}\left(1-\fr{m^2_t}{m^2_A}\right)^{3/2}+\fr{g^4}{256\pi c^4_W m^2_A}+\fr{g^4}{128\pi m^2_A}+\fr{(\la_3-\la_5)^2}{64\pi m^2_A},\ee which summarizes over annihilation channels to $tt^c$, $ZZ$, $WW$, and $HH$, respectively. Here, the approximations are given, based on the fact that $m_A$ is radically beyond the weak scale. That said, the contribution of the channel to $tt^c$ is negligible. Taking the three remaining channels into account for the relic density, we get further,
\be \langle \sigma v\rangle \simeq \left[\left(\fr{567.8\ \mathrm{GeV}}{m_A}\right)^2+\left(\fr{1.39(\la_3-\la_5)\ \mathrm{TeV}}{m_A}\right)^2 \right]\times 1\ \mathrm{pb}.\ee Here note that the correct abundance $\Om_A h^2\simeq 0.1\ \mathrm{pb}/\langle \sigma v\rangle\simeq 0.11$ requires $\langle \sigma v\rangle \simeq 1$ pb, and we use $\al=e^2/4\pi=1/128$, $g=e/s_W$, $s^2_W=0.231$. It is clear that (i) if $\la_3-\la_5\ll g^2$, the gauge portal dominates the dark matter annihilation, and the correct abundance is set for $m_A=567.8$ GeV, (ii) if $\la_3-\la_5 \gg g^2$, the Higgs portal dominates the dark matter annihilation, and the correct abundance is set for $m_A=1.39(\la_3-\la_5)$ TeV, and (iii) if $\la_3-\la_5\sim g^2$, the correct relic density is set for $m_A\sim 816$ GeV, where both the portals give significant contributions. 

As the dark scalar mass splitting $m_S-m_A\simeq \la_5 v^2/2m_A\sim 302$ MeV for $\la_5\sim 0.01$ and $m_A\sim 1$ TeV is large, the dark matter $A$ cannot scatter with nuclei via $Z$ exchange, which converts $A$ to $S$. The scattering process of $A$ with nuclei is proceeded via the Higgs portal, given at quark level via a Feynman diagram like the annihilation diagram to quarks via the Higgs portal above, inducing the effective interaction, 
\be \mathcal{L}_{\mathrm{eff}}\supset 2\la_q m_A AA \bar{q}q,\ee where $\la_q=(\la_3-\la_5) m_q/2m_A m^2_H$. Hence, the dark matter ($A$) and nucleon ($\mathcal{N}=p,n$) scattering cross section is given by \cite{bjma}
\be \sigma^{\mathrm{SI}}_{A-\mathcal{N}} =\fr{4m^2_r}{\pi}\la^2_{\mathcal{N}},\ee where the reduced mass is $m_r=m_A m_{\mathcal{N}}/(m_A + m_{\mathcal{N}})\simeq m_{\mathcal{N}}\simeq 1$ GeV, while the nucleon coupling is summed over those at quark level interactions with respective nucleon form factors to be $\la_{\mathcal{N}}/m_{\mathcal{N}} \simeq 0.35 (\la_3-\la_5)/(2m_A m^2_H)$ \cite{jelis}. Taking $m_H=125$ GeV, we have 
\be \sigma^{\mathrm{SI}}_{A-\mathcal{N}}\simeq \left[\fr{25(\la_3-\la_5)\ \mathrm{TeV}}{m_A}\right]^2\times 10^{-46}\ \mathrm{cm}^2.\ee This prediction is appropriate to the experimental limit, i.e. $\sigma^{\mathrm{SI}}_{A-\mathcal{N}} = 10^{-46}$--$10^{-45}\ \mathrm{cm}^2$ according to $m_A = 400$--1400 GeV, respectively \cite{lux}, only for the case (i) above, which requires $\la_3-\la_5\ll g^2$, thus matching $m_A=567.8\ \mathrm{GeV}=[25(\la_3-\la_5)/\sqrt{1.3}]\ \mathrm{TeV}$. Here note that $\sigma^{\mathrm{SI}}_{A-\mathcal{N}} \simeq 1.3\times 10^{-46}\ \mathrm{cm}^2$ for $m_A=567.8$ GeV \cite{lux}. It leads to $\la_3-\la_5 \simeq 0.025$, as expected. That said, the relic density is governed by the gauge portal, implying a dark matter mass $m_A=567.8$ GeV, while the direct detection cross section is set by the Higgs portal, requiring a dark matter vs. Higgs coupling $\la_3-\la_5\simeq 0.025$.\footnote{Of course, this value is appropriate to the magnitude of $\la_5\sim 0.01$ too.}         

Let us restore the family indices labeled $a,b$ in the Yukawa Lagrangian, i.e.
\be \mathcal{L}_{\mathrm{Yuk}}\supset h_{ab}\bar{l}_{aL}\eta N_{bR}+h'_{ab}\bar{l}_{aL}\eta N^c_{bL}+H.c.,\ee which gives rise to lepton flavor violation at one-loop level via processes, $e_a\to e_b\gamma$, $e_a\to 3 e_b$, and $\mu$-$e$ conversion in nuclei, where $e_a/e_b$ are assumed to be physical fields, i.e. $e,\mu,\tau$ according to $a/b=1,2,3$, respectively. Since $h'\ll h$, the contribution of the $h'$ coupling is negligible, which will be omitted. Since the current bounds for such processes indicate if $e_a\to e_b\gamma$ is constrained, the remainders are manifestly satisfied \cite{ahriche}. It is sufficient to consider only $e_a \to e_b\gamma$, derived by $\mathcal{L}_{\mathrm{Yuk}} \supset h_{ab} \bar{e}_{aL} N_{bR} H^-+H.c.$ Generalizing \cite{vicente} yields 
\be \mathrm{Br}(e_a\to e_b \gamma)=\fr{3\al v^4}{32\pi}\left|\sum_{k=1,2,3}\fr{h_{ak}h^*_{bk}}{m^2_{H^-}}F\left(\fr{M^2_k}{m^2_{H^-}}\right)\right|^2\mathrm{Br}(e_a\to e_b\nu_a\bar{\nu}_b),\ee where $N_k$ is assumed to be a physical field with mass $M_k$ by itself, without loss of generality, and $F(x)=(1-6x+3x^2+2x^3-6x^2\ln x)/6(1-x)^4$ has no pole at $x=1$, decreasing for $x$ increasing from zero, hence limited by $F(x)<1/6$ for $x>0$. The strict decay obeys 
\be \mathrm{Br}(\mu\to e\gamma)\lesssim 4.2\times 10^{-13}\left(\fr{\sum_k |h_{\mu k}h^*_{e k}|}{0.0025}\right)^2\left(\fr{770\ \mathrm{GeV}}{m_{H^-}}\right)^4,\ee with the aid of $\mathrm{Br}(\mu \to e \nu_\mu \bar{\nu}_e)\simeq 1$, $\al=1/128$, and $v=246$ GeV. This prediction is suitable to the current bound $\mathrm{Br}(\mu\to e \gamma)\simeq 4.2\times 10^{-13}$ \cite{meg}, given that $h_{\mu(e) k }\sim 0.05$ and $m_{H^-}\sim 770$~GeV, in agreement with the neutrino mass generation.
        
\section{Conclusion and outlook}

We have shown that quasi-Dirac fermions may be the source for small neutrino masses, making relevant dark matter phenomenology viable as well as comparable to charged lepton flavor violation limit. A new type of radiative inverse seesaw has been realized, providing such ingredients. It is noted that the discrete symmetry, i.e. $Z_2$, in the present setup would come from a more fundamental gauge symmetry, as justified by Krauss and Wilczek \cite{kw}. Hence, we can investigate the quasi-Dirac effect in various gauge extensions of the present model and its implication for neutrino mass and dark matter.

Indeed, a neutral vectorlike fermion is implied by various extensions of the standard model, such as dark $U(1)$ model \cite{du1,du1p}, 3-3-1-1 model \cite{3311,3311p}, trinification \cite{trini1,trini2}, etc. In the dark $U(1)$ model, a vectorlike fermion $N$ would exist due to anomaly cancellation. It may become quasi-Dirac states due to $U(1)$ breaking at low energy by a scalar field that couples to $NN$. In the 3-3-1-1 model, a vectorlike fermion may exist at the bottom of a lepton triplet for the left component $N_L$, while the right component $N_R$ is a singlet. This field obtains a Dirac mass via $SU(3)_L$ breaking, while it becomes quasi-Dirac states due to a small violation in lepton-like symmetry. The trinification contains such a vector fermion in lepton bi-triplets, which may become quasi-Dirac states by trinification breaking by a bi-sextet scalar. This work would not consider such models in detail, but the quasi-Dirac effect that translates to neutrino mass and dark matter can be easily generalized.    

\section*{Acknowledgement}

This research is funded by Vietnam National Foundation for Science and Technology Development (NAFOSTED) under grant number 103.01-2023.50. We thank Dr. Tran Van Que for a useful comment.

\end{document}